\documentclass[twocolumn,prl,showpacs]{revtex4}
\usepackage{psfrag}
\usepackage{graphicx}
\usepackage{amsmath}
\usepackage{amssymb}

\begin{document}
\title{Renormalization of Molecular Quasiparticle Levels at Metal-Molecule
  Interfaces:\\Trends Across Binding Regimes} \author{Kristian S.
  Thygesen$^1$ and Angel Rubio$^2$} \affiliation{$^1$Center for
  Atomic-scale Materials Design (CAMD), Department of Physics,
  Technical University of Denmark, DK-2800 Kgs. Lyngby, Denmark.\\
$^2$Departamento de Fisica de
  Materiales, Facultad de Qu\'{\i}micas, Centro Mixto CSIC-UPV/EHU ,
  Universidad del Pa\'{\i}s Vasco, Edificio Korta, 20018 San
  Sebasti{\'{a}}n, Spain}

\date{\today}

\begin{abstract}
  When an electron or a hole is added into an orbital of an adsorbed
  molecule the substrate electrons will rearrange in order to screen
  the added charge. This results in a reduction of the electron
  addition/removal energies as compared to the free molecule case. In
  this work we use a simple model to illustrate the universal trends
  of this renormalization mechanism as a function of the microscopic key
  parameters. Insight of both fundamental and practical importance is
  obtained by comparing GW quasiparticle energies with Hartree-Fock
  and Kohn-Sham calculations. We identify two different polarization
  mechanisms: (i) polarization of the metal (image charge formation)
  and (ii) polarization of the molecule via charge transfer across the
  interface. The importance of (i) and (ii) is found to increase with
  the metal density of states at the Fermi level and metal-molecule
  coupling strength, respectively.
\end{abstract}

 \maketitle 
The position of an adsorbed molecule's frontier orbitals with respect to the substrate
Fermi level determines the threshold energies at which electron
transfer can take place across the metal-molecule interface. Such electron transfer
processes represent a cornerstone of surface science and form the basis of photo- and non-adiabatic chemistry, organic- and molecular
electronics, as well as scanning tunneling- and
photoemission spectroscopy~\cite{gadzuk,cuniberti,nitzan03,repp05,lu04,kahn03,johnson87}.
Accurate descriptions of adsorbate energy spectra are thus 
fundamental for quantitative modeling within these important areas.

\begin{figure}[!t]
\begin{center}
\includegraphics[width=0.75\linewidth]{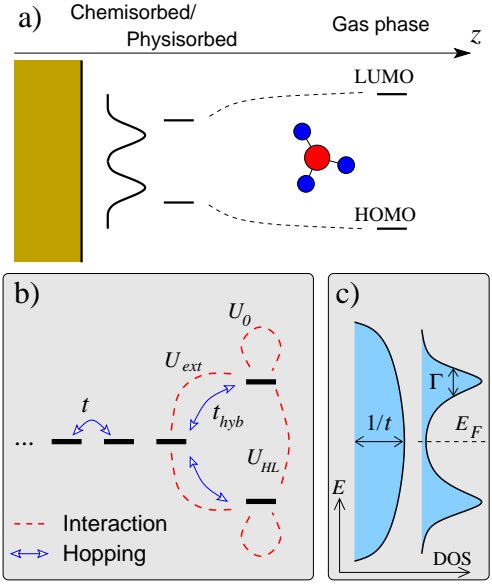}
\caption[system]{\label{fig1} (Color online) (a) Schematic of a
  molecule's HOMO and LUMO energy levels as it approaches a metal
  surface. For weak coupling (physisorbed molecule) the gap is reduced
due to image charge formation in the metal. For strong coupling
  (chemisorbed
molecule) dynamic charge transfer between molecule and metal reduces
the gap further. (b) The model used in the present study. (c) The
  semi-elliptical band at the terminal site of
the TB chain and the resonances of the molecule.}
\end{center}
\end{figure}

Recently, a number of experiments probing transport- and optical properties of molecules at metal surfaces, have found strong reductions of electron addition/removal energies due to polarization effects in the metal substrate~\cite{repp05,lu04,kahn03,johnson87,kubatkin}, see
Fig.~\ref{fig1}a. Theoretical studies of
metal-molecule interfaces are usually based on density functional theory
(DFT) and rely on an interpretation of
Kohn-Sham (KS) eigenvalues as quasiparticle energies. This interpretation is
unjustified in principle and questionable in practice -- even
with the exact exchange-correlation functional -- because
single-particle schemes like KS and Hartree-Fock (HF) take no account
of dynamical effects such as screening of added electrons/holes. Inclusion of dynamical polarization effects through classical image
charge models have been applied to correct DFT calculations of quantum
transport in molecular contacts~\cite{bda,mowbray,kristen}. Only recently, GW calculations by Neaton \emph{et al.}
have demonstrated that the energy gap of benzene is reduced by more than 3 eV when physisorbed on graphite due to image charge effects.\cite{neaton} As a complementary study, the
present work focuses on qualitative trends and explores different binding situations 
including strong-weak coupling and narrow-wide band substrates.
 
To illustrate the problem, suppose we add an electron into the
lowest
unoccupied orbital (LUMO) of an adsorbed molecule. The associated energy cost is given
by the quasiparticle (QP) spectral function
\begin{equation}\label{eq.spectral}
A_{L}(\varepsilon)=\sum_n|\langle\Psi_n^{N+1}|c^{\dagger}_{L}
|\Psi_0^N\rangle|^2\delta(\varepsilon-E_n^{N+1}+E_0^N),
\end{equation}
where $|\Psi_0^N\rangle$ and $|\Psi_n^{N+1}\rangle$ denote the
many-body $N$-particle groundstate and $(N+1)$-particle excited
states, respectively. The qualitative shape of $A_L$ is mainly dictated by the hybridization with the metal
states. On the other hand, quantitative features such as the precise position of its peak(s), depend on the system's response to the extra electron in the
LUMO. This follows by noting that peaks in $A_L$ appear at dominant Fourier components of $c^{\dagger}_L|\Psi_0^N\rangle(t)$. Dynamical effects of this type can be incorporated within an effective single-particle framework where the electron-electron interaction is represented by an energy-dependent potential known as
a self-energy. In practice the self-energy must be approximated using e.g. many-body
perturbation theory. For weak metal-molecule
coupling, total energy calculations with constrained occupations of
the molecule represent a simple alternative to the many-body approach.~\cite{gavnholt} However,
such calculations only provide a lower (upper) limit for the LUMO (HOMO) energy and, as we
show here, these limits can differ from the true quasiparticle levels.

In this Letter, we introduce a simple model of a metal-molecule
interface and use it to investigate the effect of dynamical polarization on the molecular levels as a function of microscopic key quantities such as the metal-molecule coupling strength
($\Gamma$) and the metal density of states (DOS).
We find that polarization effects consistently reduce the molecular gap as
compared to the gas phase value. This effect is completely absent in
HF theory which systematically overestimates the gap. On the other
hand, the KS spectrum is indirectly affected through the "exact density" requirement, however, the KS gap is always found to be too small. 
On the basis of GW calculations, we find that the gap reduction due to polarization of the metal, i.e. image charge
formation, is insensitive to $\Gamma$ but
increases with the metal DOS at $E_F$. On the other hand, polarization
of the molecule via charge transfer to/from the metal increases strongly
with $\Gamma$ leading to a direct correlation between adsorbate bond strength and the
gap renormalization.  

Our model Hamiltonian consists of three parts, $\hat H=\hat H_{\text{met}}+\hat
H_{\text{mol}}+\hat V$, describing the metal, the molecule, and their mutual
interaction, see Fig.~\ref{fig1}b. The metal is modeled by a non-interacting,
semi-infinite tight-binding (TB) chain,
\begin{equation}\label{eq.h_met}
\hat H_{\text{met}}=\sum_{i=-\infty}^0\sum_{\sigma=\uparrow,\downarrow} t
(c^{\dagger}_{i\sigma}c_{i-1\sigma}+c^{\dagger}_{i-1\sigma}c_{i\sigma}).
\end{equation}
The molecule is
modeled as two interacting levels representing the highest occupied (HOMO) and lowest unoccupied (LUMO) molecular orbitals
\begin{eqnarray} 
\hat H_{\text{mol}}&=&\xi_{H}\hat
n_{H}+(\xi_{H}+\Delta_0)\hat
n_{L}+\hat U_{\text{mol}}\\
\hat U_{\text{mol}} &=& U_0\hat n_{H\uparrow}\hat
n_{H\downarrow}+U_0\hat n_{L\uparrow}\hat n_{L\downarrow}+U_{HL}\hat n_{H}\hat n_{L},
\end{eqnarray}
where e.g. $\hat
n_{H}=c^{\dagger}_{H\uparrow}c_{H\uparrow}+c^{\dagger}_{H\downarrow}c_{H\downarrow}$,
is the number operator of the HOMO level. Notice that, despite the interactions, the eigenstates
of $\hat H_{\text{mol}}$ are simply Slater determinants build from
the orbitals $|H\sigma\rangle$ and $|L\sigma\rangle$.  Finally,
hybridization and interaction between the molecule and the terminal
site of the chain is described by
\begin{equation} 
\hat V=\sum_{\nu =H,L}\sum_{\sigma=\uparrow,\downarrow}t_{\text{hyb}}(c^{\dagger}_{0\sigma}c_{\nu
  \sigma}+c^{\dagger}_{\nu \sigma}c_{0\sigma})+U_{\text{ext}}\delta
  \hat n_0 \delta \hat{N}_{\text{mol}}.
\end{equation}
Here $\delta \hat n_0=(\hat n_0-1)$ and $\delta \hat{N}_{\text{mol}}=(\hat N_{\text{mol}}-2)$ represent the excess charge on
the chain's terminal site and the molecule, respectively.  We set $E_F=0$
corresponding to a half filled band, and adjust $\xi_H$ so that the
molecule holds exactly two electrons in the groundstate. Specifically
this means $\xi_H=-\Delta_0/2-U_0/2-U_{HL}$. The model neglects
interactions within the TB chain and between the molecule and interior
TB sites ($i<0$). These approximations are, however, not expected to
influence the qualitative trends described by the model.

It is instructive first to consider the model in the limit
$t_{\text{hyb}}=0$ (weak physisorption) where the metal and molecule
interact only via the Coulomb term $\hat
U_{\text{ext}}$. It is straightforward to verify that in this limit,
the many-body eigenstates coincide with the HF solutions. More
precisely they are single Slater determinants constructed from the
molecular orbitals, $|H\sigma\rangle$ and $|L\sigma\rangle$, and the
eigenstates of the Hamiltonian, $\hat
H_{\text{met}}^{\text{HF}}(\delta N_{\text{mol}})=\hat
H_{\text{met}}+U_{\text{ext}}\langle\delta
\hat{N}_{\text{mol}}\rangle\hat n_0$, where $\langle\delta
\hat{N}_{\text{mol}}\rangle=-2,\ldots,2$ is the number of molecular
orbitals in the Slater determinant.  Although the eigenstates are single
Slater determinants, the situation is different from the
non-interacting case because the single-particle orbitals of the metal
depend on the occupation of the molecule (each electron on the
molecule will shift the potential of the terminal site by
$U_{\text{ext}}$). As we will see below this has important
consequences for the position of the molecule's QP levels. 

\begin{figure}[t]
\begin{center}
\includegraphics[width=0.95\linewidth]{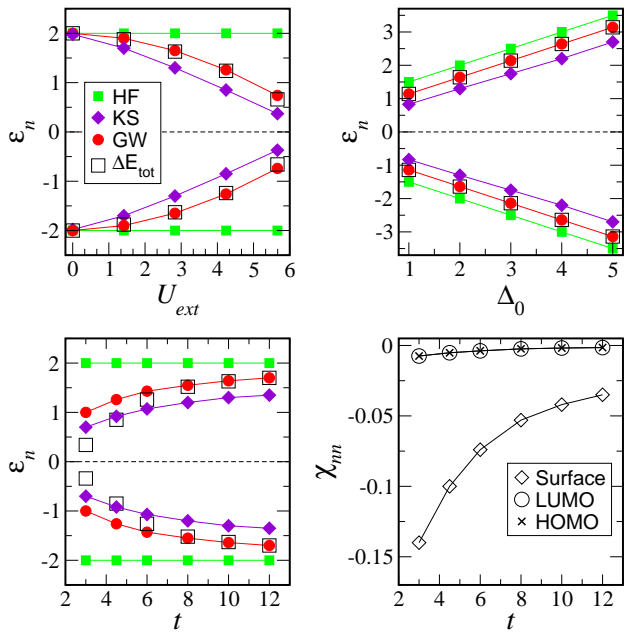}
\caption[system]{\label{fig2} (Color online) Position of the HOMO and
  LUMO levels as a function of different model parameters for a weakly
  coupled molecule (small $\Gamma$). Note that dynamical polarization is completely absent
  in HF theory and incorrectly described by the ``exact'' KS theory. 
Open squares denote the exact total energy difference between
  the normal groundstate and the groundstate when the molecule has been constrained 
to hold one extra electron/hole. Lower right panel shows on-site elements of the
  static, linear response function of the HOMO, LUMO and terminal
site of the TB chain. Clearly, the renormalization of the gap is due
to polarization of the metal.}
\end{center}
\end{figure}

In order to test the accuracy of the GW approximation for the problem
at hand, we have compared it to exact diagonalization results in the
case where the TB chain is truncated after the first two sites, see the
supplementary material.~\cite{EPAPS} The GW
spectral function is essentially identical to the exact result, also
for strong interactions, demonstrating that GW captures
the essential physics of the model accurately. 

To study how the molecular levels depend on the model parameters, we
vary one parameter at a time keeping the remaining fixed at the
following reference values: $t=10$, $t_{\text{hyb}}=0.4$, $U_0=4$,
$U_{HL}=3$, $U_{\text{ext}}=2.8$, $\Delta_0=2$. The reference values
correspond to weak coupling ($\Gamma\approx 0.02$) and a wide-band
metal ($W=4t=40$).  The positions of the HOMO and LUMO levels are defined as the maximum of
the corresponding spectral functions (\ref{eq.spectral}) which we calculate from the Green's function,
$A_{n}(\varepsilon)=-(1/\pi) \text{Im}G^r_{nn}(\varepsilon)$, see
supplementary material for plots of a representative set of spectral
functions~\cite{EPAPS}. The Green's
function is obtained by solving the Dyson equation fully
self-consistently in conjunction with the self-energy
($\Sigma_{\text{GW}}[G]$ or $\Sigma_{\text{HF}}[G]$) as described in
Ref.~\cite{gw_prb}. Following the
lattice version of DFT~\cite{schonhammer95}, we define the KS Hamiltonian as the non-interacting part of $\hat H$ with the on-site energies corrected to yield the exact occupation numbers. Using the GW occupations as "exact" target occupations this allows us to obtain the "exact" KS levels.

In the upper left panel of Fig.~\ref{fig2} we show the position of the molecule's QP
levels as function of the metal-molecule interaction,
$U_{\text{ext}}$. The GW gap decreases as $U_{\text{ext}}^2$. The
reduction of the gap corresponds to the energy gained
by letting the HF states of the metal relax in response to the added
electron/hole, i.e. to the perturbations $\pm U_{\text{ext}}\hat n_0$.
Indeed, the open squares show the difference in total energy
between the normal groundstate and the groundstate when the
molecule has been constrained to hold one extra
electron/hole.\cite{note3} HF completely misses this effect due to its neglect of orbital relaxations, and consequently the HF gap
becomes too large. In a screening picture, the difference between the HF and
GW levels equals the binding energy between the added electron/hole
and its image charge (the positive/negative induced density at the
terminal site), corrected by the cost of forming the image charge. 

The "exact" KS theory underestimates the gap, but seems to reproduce the
trend of the GW calculation. At first this is surprising since a mean-field
theory cannot describe the dynamical effects responsible for the gap
reduction. The explanation is that the KS levels are affected \emph{indirectly}: the KS levels are forced to follow the GW levels in order to reproduce the GW occupations. 

In the upper right panel of Fig.~\ref{fig2} we show the dependence of
the QP levels on the molecule's intrinsic gap, $\Delta_0$. The fact that the
renormalization of the QP gap due to image charge formation is independent of $\Delta_0$ follows from the discussion above since $\hat H_{\text{met}}^{\text{HF}}$ is independent of
$\Delta_0$. 

It seems intuitively clear that the size of the gap reduction (for fixed $U_{\text{ext}}$) should depend on the
polarizability of the metal, i.e. how much the electron density
changes in response to the perturbing field created by the added electron/hole.
In the lower left panel of Fig.~\ref{fig2} we show the
QP levels as function of the chain hopping, $t$.\cite{note4} The gap reduction is larger for small $t$ corresponding to a narrow
band. This is easily understood by noting that a narrow band implies a large
DOS at $E_F$ which in turn implies a larger density response
function. In the lower right panel of Fig.~\ref{fig2} we show the
diagonal elements of the static component of the response
function, $\chi_{nn}(\omega=0)$, for the HOMO,
LUMO and terminal site of the chain, respectively. In this weakly coupled regime,
the response function of the HOMO
and LUMO is clearly negligible for all values of $t$, while the response of
the terminal site is significant and increases as $t$ is
reduced. This clearly demonstrates that the QP levels are renormalized
by screening inside the metal. In general, only the response at frequencies $|\omega| \leq \Gamma$ is relevant as $\Gamma^{-1}$ sets the decay time of
the states $c^{\dagger}_L|\Psi_0\rangle$ and $c_H|\Psi_0\rangle$.

Interestingly, the deviation between the GW levels
and $\Delta E_{\text{tot}}$ becomes significant for small $t$. We stress that
this does not imply that the GW results are wrong. In fact, $\Delta
E_{\text{tot}}$ only represents an upper/lower bound for the true QP energies, and
the deviation thus indicates that the overlap of
$c^{\dagger}_L|\Psi_0\rangle$ and $c_H|\Psi_0\rangle$ with higher
lying excited states of the $(N+1)$-particle system (see
Eq. (\ref{eq.spectral})) is more important for small $t$. 

\begin{figure}[t]
\begin{center}
\includegraphics[width=0.95\linewidth]{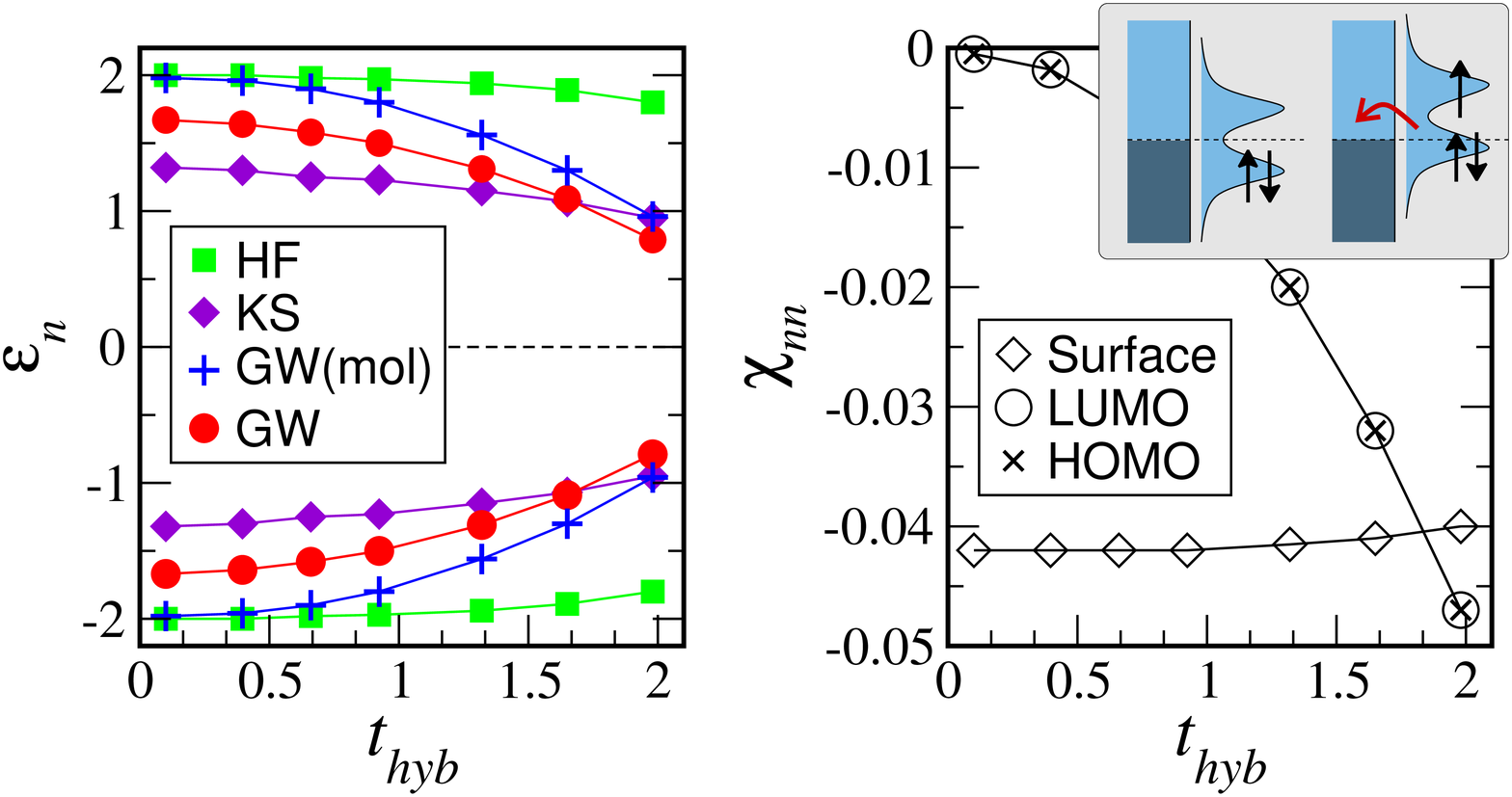}
\caption[system]{\label{fig3} (Color online) Left: Position of
  the molecule's HOMO and LUMO levels as a function of the hybridization strength, $t_{\text{hyb}}$. GW(mol) refers to a calculation where only the
  internal interactions on the molecule, $\hat U_{\text{mol}}$, have been treated within
  GW while the interactions with the metal, $\hat U_{\text{ext}}$ have been
  treated within HF. Right: Static response function for
  the HOMO, LUMO, and terminal site of the TB chain. The cartoon
illustrates how dynamic charge transfer stabilizes the
charged system and thereby reduces the gap.}
\end{center}
\end{figure}

We now leave the regime of weak molecule-metal coupling and consider
the dependence of the molecular spectrum on $t_{\text{hyb}}$, see
Fig.~\ref{fig3}. At finite $t_{\text{hyb}}$ the molecular levels
broaden into resonances, however, it is still possible to define the
level position as the resonance maximum.\cite{width_note} 
In addition
to the GW, HF, and KS levels, we also show the result of a 
calculation where only $\hat U_{\text{mol}}$ is
treated at the GW level while $\hat U_{\text{ext}}$ is treated at
the HF level. This means that the only dynamical effects included in
the GW(mol) calculation are those due to $\hat
U_{\text{mol}}$. In particular, image charge formation in the metal is ignored.
 
Focusing on the GW and GW(mol) results, we conclude that the reduction of the QP gap as function of $t_{\text{hyb}}$
is mainly due to the interactions internally on the molecule. On the other hand, the reduction due to the molecule-metal
interaction is largely insensitive to
$t_{\text{hyb}}$ (this reduction is given by the difference between the
GW and GW(mol) levels). From the right panel of Fig.~\ref{fig3}, we
see that $|\chi_{nn}|$ for the HOMO and LUMO states increases
with $t_{\text{hyb}}$ indicating that the gap reduction due to $\hat U_{\text{mol}}$ is of a
similar nature as the image charge effect, but with the molecule
itself being polarized instead of the metal. Polarization of the
molecule can occur via dynamic charge transfer to/from the metal as
illustrated in the cartoon of Fig.~\ref{fig3}. Note that this picture is consistent
with the fact that no polarization of the molecule occurs without
coupling to the metal (see GW(mol) result in the limit $t_{\text{hyb}}\to 0$ in Fig.
\ref{fig3}). 

The fact that dynamical polarization becomes more important as the molecular DOS at the
Fermi level increases, has important consequences for charge transport
in molecular junctions. Indeed, as the chemical potential,
$\mu_{\alpha}$, approaches a molecular level, polarization effects become stronger and the level is shifted towards $\mu_{\alpha}$. This effect
was recently shown to have a large impact on the junction $IV$
characteristics.~\cite{gw_prl} Finally, we mention that the
correlation between chemisorption bond strength and gap
renormalization suggested by Fig.~\ref{fig3}, has in fact been
observed in inverse photoemission spectroscopy
experiments.\cite{johnson87}

In summary, we have used a simple model of a metal-molecule interface to identify universal trends in the way dynamical polarization renormalize molecular interface states. The effect of polarization is to reduce the gap between occupied and unoccupied molecular orbitals. The size of the gap reduction  
correlates directly with the static response function, and is promoted by
larger metal and/or molecular DOS at the Fermi level. The latter
suggests that transition metals with half-filled $d$ band should be
more effective in reducing the gap of molecular adsorbates than
e.g. alkali, sp, and the early/late
transition metals. The results strongly indicate that dynamical polarization is of fundamental importance for charge transfer at metal-molecule interfaces.

The authors thank K. Flensberg and K. Kaasbjerg for stimulating discussions.
KST acknowledges support from the Danish Center for
Scientific Computing and The Lundbeck
Foundation's Center for Atomic-scale
Materials Design, and AR from MEC (FIS2007-65702-C02-01), Grupos Consolidados
UPV/EHU  (IT-319-07); EU e-I3 ETSF project.



\begin{thebibliography}{30}
\bibitem{note3} In the
        relevant limit $t_{\text{hyb}}=0$, $\Delta E_{\text{tot}}$ can be obtained from the change in total energy upon shifting the on-site energy of the terminal site of a semi-infinite TB chain by
        $\pm U_{\text{ext}}$, which in turn can be accurately calculated .
\bibitem{note4} We stress that even though $\Gamma$ depends on $t$,
  all results of Fig. \ref{fig2} are essentially unchanged as $t_{\text{hyb}}$
  is reduced below the reference value and thus represent the
  limit $t_{\text{hyb}}\to 0$.
\bibitem{width_note}
Due to the interactions, the resonance width is slightly enlarged
relative to the non-interacting value of
$\Gamma^{-1}\approx|t|/|t_{\text{hyb}}|^2$.
\end{thebibliography}
\end{document}